\documentstyle[epsf,12pt]{article}
{\end{list}}
\newcounter{enumct}

\newcommand{\captive}[1]{\rule{5mm}{0mm}%
\begin{minipage}{150mm}\caption[small]{#1}\end{minipage}}

\newcommand{\be}{\begin{equation}}
\newcommand{\ee}{\end{equation}}
 \newcommand{\ovl}{\overline}
\newcommand{\bea}{\begin{eqnarray}}
\newcommand{\eea}{\end{eqnarray}}
\newcommand{\ba}{\begin{array}}
\newcommand{\ea}{\end{array}}
\newcommand{\beqa}{\begin{eqnarray}}
\newcommand{\eeqa}{\end{eqnarray}}

\newcommand{\lsim}{{\;\raise0.3ex\hbox{$<$\kern-0.75em\raise-1.1ex
\hbox{$\sim$}}\;}}
\newcommand{\gsim}{{\;\raise0.3ex\hbox{$>$\kern-0.75em\raise-1.1ex
\hbox{$\sim$}}\;}}

\newcommand{\PL}[1]{Phys. Lett.\ { #1}}

\newcommand{\PRL}[1]{Phys. Rev. Lett.\ { #1}}

\newcommand{\ssu}{$SU(2)_L\times SU(2)_R\times U(1)_{B-L}\,$}
\newcommand{\sul}{$SU(2)_L$}
\newcommand{\sulu}{$SU(2)_L\times U(1)_Y$}
\newcommand{\sur}{$SU(2)_R$}
\newcommand{\matr}{\left( \begin{array}}
\newcommand{\ematr}{\end{array} \right)}

\newcommand{\dis}{\displaystyle}

\newcommand{\lr}{{left-right symmetric model}}

\begin{document}
 
\mbox{}\vspace*{-1cm}\hspace*{9cm}\makebox[7cm][r]{  HIP-1999-77/TH}
\medskip
 
\Large
 
\begin{center} {\bf Tests of  the left-right  electroweak
model at linear collider}
 
\normalsize
{K. Huitu,}$^a$ J. Maalampi,$^b$ 
{P. N. Pandita,}$^c$
 { K. Puolam\"aki}$^a$, \mbox{ M. Raidal}$^{d}$ and  \mbox{N. Romanenko}$^{b,e}$ 
\\[15pt]$^a${\it Helsinki Institute of
Physics, Helsinki, Finland}\\$^b${\it Department of Physics, Theoretical Physics Division,
University of Helsinki, Finland}\\
$^c$ {\it Department  of Physics, North Eastern Hill University, Shillong, India}\\
$^d${\it 
DESY, Germany}\\
$^e$ {\it  Petersburg Nuclear Physics Institute,
   Gatchina, Russia}\\[15pt]

\bigskip
 
\normalsize
 
{\bf\normalsize \bf Abstract}

\end{center}
 
\normalsize

The left-right model  is a gauge theory of electroweak interactions based on the
gauge symmetry  \ssu . The main motivations for this model are that it gives an
explanation for the parity violation of weak interactions, provides a mechanism
(see-saw) for generating neutrino masses, and has $B-L$ as a gauge symmetry.
The quark-lepton symmetry in weak interactions is also maintained in this theory. The
model has many predictions one can directly test at a TeV-scale linear collider. We will consider here two processes ($e^-e^- \rightarrow 
q\:q\:\bar q\:\bar q$  and
 $e^-e^- \rightarrow
\mu\:\nu\: q\:\bar q \; $)
 testing the lepton flavour violation  predicted by the model. We will also discuss constraints on supersymmetric versions of the model.\bigskip


 \normalsize
 
\section{Introduction} 

The left-right symmetric model (LRM) has the gauge symmetry \ssu, which is spontaneously
broken to the Standard Model (SM) symmetry \sulu\  at low energies \cite{lrm}. The
Lagrangian of the model is   parity conserving in the symmetric limit. The parity
violation  of weak interactions observed at low-energy domain is generated dynamically
via the spontaneous  breaking of the gauge symmetry. This is in contrast with
the SM which is parity violating by definition, and it was originally the main motivation
for the LRM.  

 The LRM  differs from the Standard  Model (SM) also in another essential respect.
Yukawa couplings between neutrinos and the fundamental scalars   give rise to
the see-saw mechanism \cite{seesaw}, which provides the simplest explanation for the
lightness of  neutrinos, if neutrinos do have a mass. In the SM neutrinos are exactly
massless by  construction. The recent observations on the  atmospheric
 neutrino fluxes by the Super-Kamiokande \cite{atmos} seem to verify that neutrinos indeed  have a
 mass. Massive neutrinos are also indicated by the observed deficit of solar neutrinos \cite{sun} and   the 
observation of
$\bar\nu_{\mu}\to\bar\nu_{e}$ transitions in the LSND experiment \cite{LSND}. 

So far there has been, however,  no  evidence of  left-right symmetry  in
weak interactions but everything seems to be well described in terms of the ordinary
$V-A$ currents.  This fact can be used
 to set constraints on the parameters of the LRM, such as the
masses of  the new gauge bosons
$W_R^{\pm}$ and 
$Z_R$,  associated with the gauge symmetry
 \sur, and their  mixings with the \sul\  bosons
$W_L^{\pm}$ and  $Z_L$ \cite{gunion,phen,deshpande}. Such constraints
depend quite crucially on the assumptions one makes. This is true, for example, for 
the often quoted mass
bound $M_{W_R}\gsim 1.6$ TeV from the $K_S-K_L$ mass difference \cite{BBS},
as well as for the limit $M_{W_R}\gsim 1.1$ TeV from the double beta decay \cite{Hirsch}.

In general, the new weak gauge bosons $W_R$ and $Z_R$ mix with their SM counterparts, but the observed $V-A$ structure of the weak force indicates that this mixing is quite small. Let us denote 
 the heavier mass eigenstate charged boson as the superposition
$W_2=\sin\zeta W_L + \cos\zeta W_R$ and identify the orthogonal state $W_1=\cos\zeta W_L
-\sin\zeta W_R$ with the ordinary $W$ boson. In the case of the manifest
left-right symmetry semileptonic-decay data can  be used to derive an upper limit 
of 0.005 on the mixing angle $\zeta$ \cite{Wolfenstein}.
 From neutral current data one can derive the lower bound 
$M_{Z_2}\gsim 400$ GeV for the mass of the new
$Z$ boson and the upper bound of
$0.008$ for  the $Z_1,Z_2$ mixing  angle \cite{phen}, where $Z_1$ denotes the ordinary neutral weak boson. 

At the Tevatron  direct searches have been made for
$W_2$ ($\simeq W_R$) in the channels $pp\to W_2\to eN\; {\rm or}\; \mu N$, where $N$ is a
neutrino. In the LRM $N$ should be identified with the heavy right-handed neutrino, since
the coupling of the ordinary light neutrino to $W_2$ is strongly suppressed.
The most stringent bound announced is 
$M_{W_2}\gsim 720$ GeV \cite{D0mass}. It should be emphasized that   this bound is based on
several assumptions. It is assumed that the quark-$W_2$ coupling has the SM
strength, the CKM matrices $V_L^{\rm CKM}$ and $V_R^{\rm CKM}$ are equal, and the
right-handed neutrino does not decay in the detector but appears as missing $E_T$. It has
been argued that if one relaxes the  first two assumptions, the mass bound will be 
degraded considerably
\cite{rizzoap95}. 

The Tevatron mass limit for the new neutral intermediate boson from the dimuon and
dielectron decay channels is
$M_{Z_2}\gsim 620$ GeV \cite{Zmass}.

We have considered the linear collider \cite{PRep} phenomenology of the left-right symmetric model, both with and without supersymmetry, 
in several previous publications \cite{We}, and we summarized the results of our studies in the foregoing issue of this series \cite{123E}. Here we shall report the studies we have carried out  since that previous summary.
In Section 2 we briefly recall the basic features of the LRM. In Section 3 we will consider various processes where one can test the lepton number violation predicted by the model.  The supersymmetric left-right model (SLRM) is described in Section 4
and we discuss various theoretical and phenomenological constraints on it. Section 6 is a
summary.

\section{Description of the left-right symmetric model}

In the left-right symmetric model quark and leptons are assigned to the doublets of
the gauge groups $SU_{L}(2)$ and $SU_{R}(2)$ according to their chirality \cite{Rabibook}:

\bea
 &\Psi_L = {\matr{c} \nu_e \\ e^-\ematr_L} = (2 ,1,-1),\;\;
&\Psi_R = {\matr{c} \nu_e \\ e^-\ematr_R} =
(1,2 ,-1),\nonumber\\
&Q_L = {\matr{c} u \\ d\ematr_L}=\left( 2 , 1 ,{\Large \frac13} \right),\;\;
&Q_R = {\matr{c} u \\ d\ematr_R} = \left( 1, 2 ,{\normalsize\frac13} \right),
\eea
 and similarly for the other families. The minimal set of fundamental scalars, the
theory to be symmetric under the $L\leftrightarrow R$ transformation, consists of the
 following Higgs multiplets:
\be
\begin{array}{c} {\dis\Phi =\matr{cc}\phi_1^0&\phi_1^+\\\phi_2^-&\phi_2^0
\ematr = (2,2,0)},
\\[15pt]
 {\dis\Delta_L =\matr{cc}\Delta_L^+&\sqrt{2}\Delta_L^{++}\\
\sqrt{2}\Delta_L^0&-\Delta_L^+
\ematr = ({ 3},{ 1},2)},\\[15pt]
{\dis\Delta_R=\matr{cc}\Delta_R^+&\sqrt{2}\Delta_R^{++}\\
\sqrt{2}\Delta_R^0&-\Delta_R^+\ematr = ({ 1},{ 3},2)}.
\end{array}
\ee
They transform according to $\dis\Phi\to U_L\dis\Phi U_R^{\dagger}$, $\Delta_L
\to U_L\Delta_L U_L^{\dagger}$ and $\Delta_R
\to U_R\Delta_R U_R^{\dagger}$, where $U_{L(R)}$ is an element of  $SU(2)_{L(R)}$.
The vacuum expectation value of the bidoublet $\dis\Phi$ is given by
\be
\begin{array}{c}
{\dis\langle\Phi\rangle=\frac1{\sqrt{2}}\matr{cc}\kappa_1&0\\0&\kappa_2\ematr.}
\end{array}\label{kappa}
\ee 
This breaks the Standard Model symmetry \sulu, and it  generates masses
to fermions through the Yukawa couplings
\be 
{\cal L}^{\rm Yukawa}_{\Phi}= \bar\Psi_L^i(f_{ij}\Phi +
\tilde f_{ij}\tilde\Phi)\Psi_R^j +  \bar Q_L^i(f_{ij}\Phi +
\tilde f_{ij}\tilde\Phi)Q_R^j + h.c.,
\ee
 where $\tilde\Phi=\sigma_2\Phi^*\sigma_2$. 

The vacuum expectation values of the scalar triplets 
are denoted by

\be
\begin{array}{c}  {\dis\langle\Delta_{L,R}\rangle
=\frac1{\sqrt{2}}\matr{cc}0&0\\v_{L,R}&0
\ematr.}
\end{array}\label{vlr}
\ee
The right-triplet $\dis\Delta_R$ breaks the $SU(2)_R\times U(1)_{B-L}$ symmetry,
and at the same time the discrete $L\leftrightarrow R$ symmetry, and it yields a
Majorana mass to the right-handed neutrinos through the Yukawa coupling
\be
  {\cal{L}}^{\rm Yukawa}_{\Delta} = ih_L \Psi_L^TC\sigma_2\Delta_L\Psi_L +ih_R
\Psi_R^TC\sigma_2\Delta_R\Psi_R + h.c.,
\label{yukawa}
\ee
 where $\Delta_{L,R}=\Delta_{L,R}^i\sigma_i$. The  conservation of electric charge
prevents the triplet Higgses from coupling to quarks. The Yukawa Lagrangian (\ref{yukawa}) is the origin of lepton number violating interactions,
a novel feature of the LRM.

 In order to have the tree-level value of the $\rho$
parameter  close to unity,
   $\rho = 0.9998 \pm 0.0008$ \cite{Caso}, one should assume
$v_L\lsim 9$ GeV. 

 It is argued in  ref.
\cite{deshpande} that to suppress the FCNC one must require  the Higgs potential
to be such that in the minimum $\kappa_1\ll \kappa_2$ or $\kappa_1\gg
\kappa_2$. This requirement has the consequence that the $W_L,W_R$ mixing angle
$\zeta$ is necessarily small, since $\zeta\sim (g_L/g_R)^2|\kappa_1\kappa_2|
/|v_R|^2$. 

In the general case, the charged current Lagrangian  is given by
\begin{equation}
\begin{array}{lcr}
{\cal L}^{CC}_{wk}&\!\simeq\!&\frac{\displaystyle g_L}{\displaystyle 2\sqrt{2}}
\left[{\left({\cos\zeta\,{J^+_L}_\mu+ \sin\zeta\,
{J^+_R}_\mu}\right){W^+_1}^\mu}\right.\;\;\;\;\;\;\;\;\;
\;\;\;\;\;\;\;\;\;\;\;\;\;\;\;\;\;\;\;\;\;\;\;\;\;\;\;\;\;\;\;\;\;\;\;\;\;\;\\
&&+\left.{ 
\left({\cos\zeta\,{J^+_R}_\mu} - \sin\zeta\,
{J^+_L}_\mu\right){W^+_2}^\mu+ {\rm h.c.}}\right].
\;\;\;\;\;\;\;\;\;\;\;\;\;\;\;\;\;\;\;\;\;\;\;\;\;\;\;\;\;\;\;\; \\
\end{array}\label{HeffCC}
\end{equation}
where 
$J^+_{R/L}=\overline e\gamma(1\pm \gamma_5)\nu_{R/L}$. In the limit $\zeta\to 0$,
the couplings of
$W_1\simeq W_L$ are similar to those of the  $W$ boson in the SM. We have assumed that the gauge couplings $g_L$ and $g_R$ associated with  the symmetries $SU(2)_L$
and $SU(2)_R$, respectively, are equal, as the manifest left-right symmetry would necessitate.

The  neutral current Lagrangian is given in terms of mass 
eigenstate bosons 
by
\begin{equation}
\begin{array}{lcr}
{\cal L}^{NC}_{wk}&\!=\!&eJ^{em}_\mu A^\mu+
\frac{\displaystyle g_L}{\displaystyle\cos\theta_W}
\left\{{\left[{{J^Z_L}_\mu\!-\lambda\cos\theta_W\!\left({\sin^2
\theta_W{J^Z_L}_\mu
\!+\cos^2\theta_W{J^Z_R}_\mu}\right)}\right]\!Z_L^\mu}\right.\;\;\;\;\\
& &\!+\left.{\left({\cos^2\theta_W}\right)^{-1/2}\left({\sin^2\theta_W
{J^Z_L}_\mu\!+\cos^2\theta_W{J^Z_R}_\mu }\right)\!Z_R^\mu }\right\}\, ,
\end{array}\label{HeffNC}
\end{equation}
where $\lambda= (M_{W_L}/M_{W_R})^2$, and 
the weak neutral currents are given by $J_{L/R}^Z=J_{L/R}^3-Q\sin^2
\theta_W\, J^{em}$ with  $J_{L/R}=\overline\psi\gamma T_{L,R}^3\psi$ and
$J^{em}=\overline\psi\gamma Q\psi$. The couplings of $Z_L$ approach the SM $Z$
couplings in the limit $\lambda\ll 1$.

{}From the left-handed and right-handed neutrino states one  can form three
types of Lorentz-invariant mass terms: Dirac term $\ovl\nu_L\nu_R$, and
Majorana terms
$\ovl\nu^c_L \nu_R$ and  $\ovl\nu^c_R\nu_L$, where the latter two terms break the
lepton number by two units. All these terms are realized in the
\lr\ with the Yukawa coupling (\ref{yukawa}) and the vevs given in eqs.
(\ref{kappa}) and (\ref{vlr}). 
The  see-saw mass matrix of neutrinos is given by 
\be
   M = \left(   \begin{array}{cc} m_L & m_D \\ m_D^T & m_R 
\end{array} 
\right).
\label{seesaw}\ee
 The entries are $3\times 3$ matrices given by $m_D=(f\kappa_1
+ g\kappa_2)/\sqrt{2}$,
$m_L = h_L v_L$ and
$m_R=h_R v_R$. The mass of the charged lepton is given by
$m_l=(f\kappa_2 + g\kappa_1)/\sqrt{2}$, and therefore if $f$
and
$g$ are comparable, one has $m_D\simeq m_l$. Unless there is an
extraordinary hierarchy among the couplings, one has
$m_L\ll m_D\ll m_R$. In this case the approximate masses of
the  Majorana states that diagonalize the neutrino Lagrangian
are given by 
  $ m_{\nu} \simeq m_D^Tm_R^{-1}m_D$ and $m_{N} \simeq 
m_R$.

\bigbreak

\section{Some tests of the LRM in  $e^-e^-$-collisions. }

\newcommand{\lbl}[1]{\label{eq:#1}}
\newcommand{\procq}{ $e^-e^- \rightarrow 
q\:q\:\bar q\:\bar q$ }
\newcommand{\prmnqq}{ $e^-e^- \rightarrow
\mu\:\nu\: q\:\bar q \; $  }
\newcommand{\prmnud}{ $e^-e^- \rightarrow
\mu \:\nu \: d \:\bar u \; $ }
\newcommand{\Del}{\Delta_L^-}
\newcommand{\DDel}{\Delta_L^{--}}
\newcommand{\bi}{\bibitem}

 The main prediction of left-right symmetric model to be tested at future experiments is the
existence of the
right-handed gauge bosons $W_R$ and $Z_R$. Another essential prediction  of the model are the  Higgs triplets. In what follows we will show that the process \procq 
provides a good place to test the right-handed gauge sector below the $W_R$ threshold, while
the process \prmnqq is very promising for the search of the left-handed  triplet Higgses $\Delta_L$.

\subsection{Process \procq as a test of the LRM below $W_R$ threshold.}

Let us first give the arguments that make us to consider the reaction \procq particularly suitable for testing  the LRM. First of all, the final state particles are all light, so that there is no kinematical suppression for the process, in contrast with, e.g., the $W_R$ pair production. Consequently, one may expect 
to detect evidence of the LRM through this reaction well below the  $W_R$ threshold.
 Reactions with ordinary neutrinos in the final state are not very useful as invisibility of neutrinos makes them not easy to distinguish from the background processes. Also, reactions with final state electrons are not 
that good because of the possible mix-up of the initial and final state particles.

One could consider, of course, for the leptonic final states, for example for the reaction
$e^-e^- \rightarrow \mu^-\:\mu^-\:\mu^-\:\mu^+$.
 The  reactions like $e^-e^- \rightarrow
b\:b\:\bar{t}\:\bar{t} $ that  involve charged currents but not neutral currents offer
 however a more unambiguous test  of the LRM than the leptonic processes. 
We prefer final states with
  $b$-quarks as the $b$-jets are relatively 
  easy to identify in experiment 
    \cite{redbook}.
 From this point of view,
 the best  process for a study would be $e^-e^- \rightarrow
b\:b\:\bar{t}\:\bar{t} $. However, as  will be seen from our numerical results,
it will possible to measure the cross section also  for the  4-jet reactions with no
$b$-jets, as well as  for the reactions with a single $b$-jet.

 We have derived  the squared matrix elements for $e^-e^- \rightarrow b\:b\:\bar{t}\:\bar{t} $ and computed the ensuing  cross sections  at the collision energies
$\sqrt s=1$ TeV and $\sqrt s=1.5$ TeV  by means of CompHEP \cite{CompHEP}.

In Fig. 1 we show the energy dependence
of the total cross section of the process $e^-e^- \rightarrow
b\:b\:\bar{t}\:\bar{t} $  for various values of
masses of the right-handed triplet Higgs $\Delta_R^{--}$ and the 
right-handed neutrino $\nu_2$. In all the cases
the right-handed boson mass is taken to be $M_{W_R}= 700$ GeV.

\begin{figure}
\vspace*{-0.5cm} \hspace*{0.2cm}
\epsfysize=10cm
\epsfxsize=14cm
\epsffile{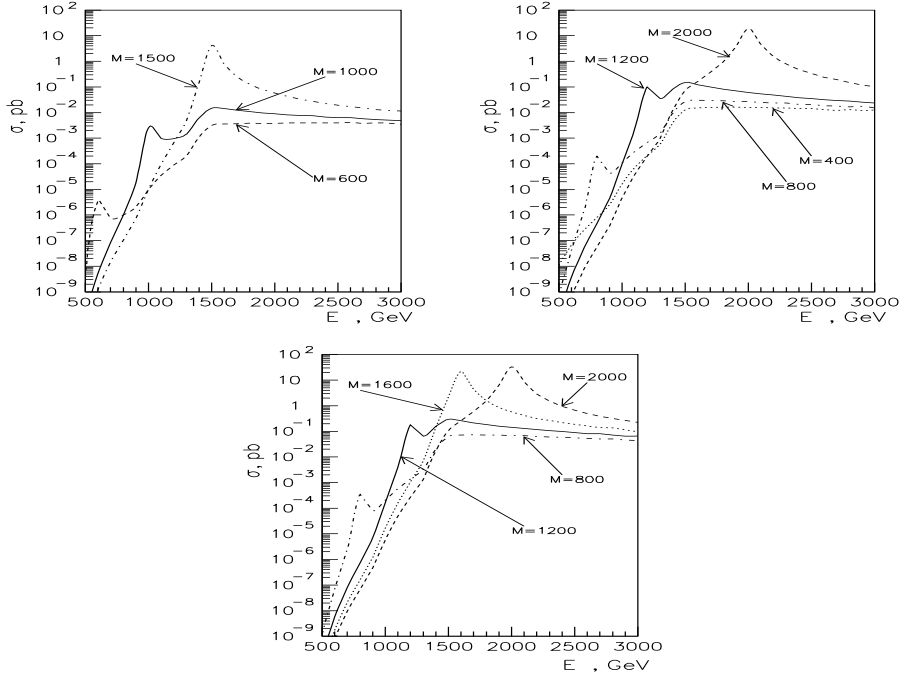}
\caption{\footnotesize Energy dependence
of the full cross section for the process
$e^-e^- \rightarrow b\:b\: \bar{t}\: \bar{t}$ 
  for  different values
 of $\Delta_R^{--}$ mass
 ($M \equiv M_{\Delta_R^{--}}$)
 and right-handed neutrino masses:
 $m_{\nu_2}=1$ TeV (left upper picture),
 $m_{\nu_2}=1.5$ TeV (right upper picture),
 $m_{\nu_2}=2$ TeV (lower picture).}
\label{fe}
\end{figure}

 In Fig. 2
we present the cross section and
 sensitivity contours
for $\sqrt s=1.5$ TeV with the masses of the right-handed
neutrinos  equal to 1.5 TeV for the abovementioned process and for the
process with 1 b-jet (or  only with light quarks)
in the final state.
The achievable limit for
$M_{W_R}$ is now about 1.5 TeV at the triplet Higgs
resonance and outside the resonance  about 1 TeV,
 a considerable improvement to the present bound.
As the cross section is proportional to the mass of neutrino, the larger
$m_{\nu_2}$ the more stringent are the ensuing constraints.
Following the arguments of \cite{redbook}
we apply the following cuts:
each b-jet should have energy more than 10 GeV;
 each t-jet should have 
energy more than 190 GeV;
 the opening 
angle between two detected jets should be
greater than $20^{\circ}$;
the angle between each detected jet and
the colliding axis should be greater than 
$36^{\circ}$;
the total energy of the event should be
greater than 400 GeV.
We have tested that when these cuts are imposed
 the following relations hold between the cross sections of the reactions with no, one and two $b$-jets in the final state:   
   
   \be
   \sigma(0b)\approx\sigma(1b)\approx 4 \cdot \sigma(2b); \:
   \lbl{rel}
   \ee
These relations may be very useful as a test
of the LRM.

\begin{figure}[t]
\vspace*{-0.5cm} \hspace*{0.2cm}
\epsfysize=13cm \epsfxsize=6.5cm \epsffile{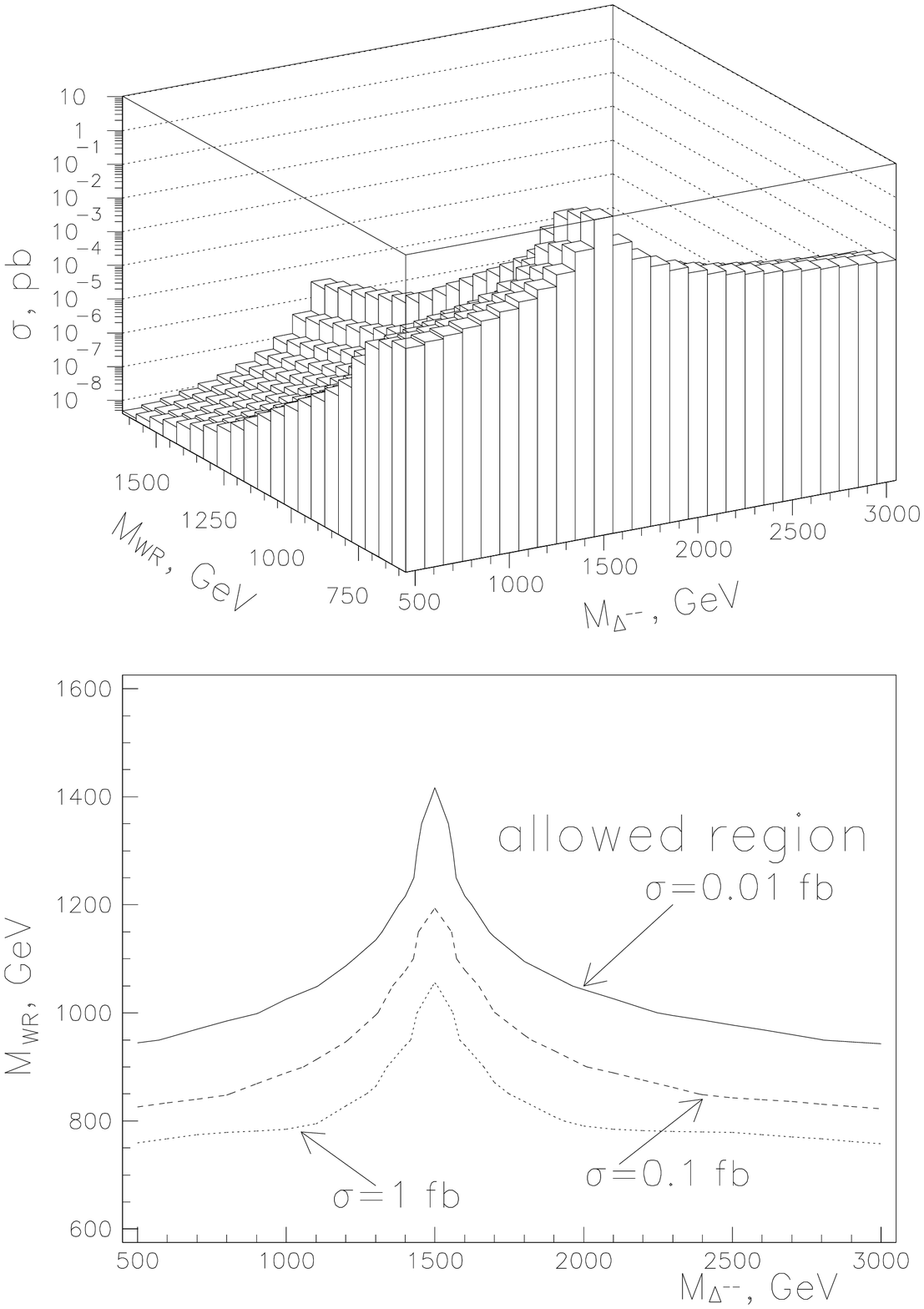}
\epsfysize =13cm \epsfxsize=6.5cm \epsffile{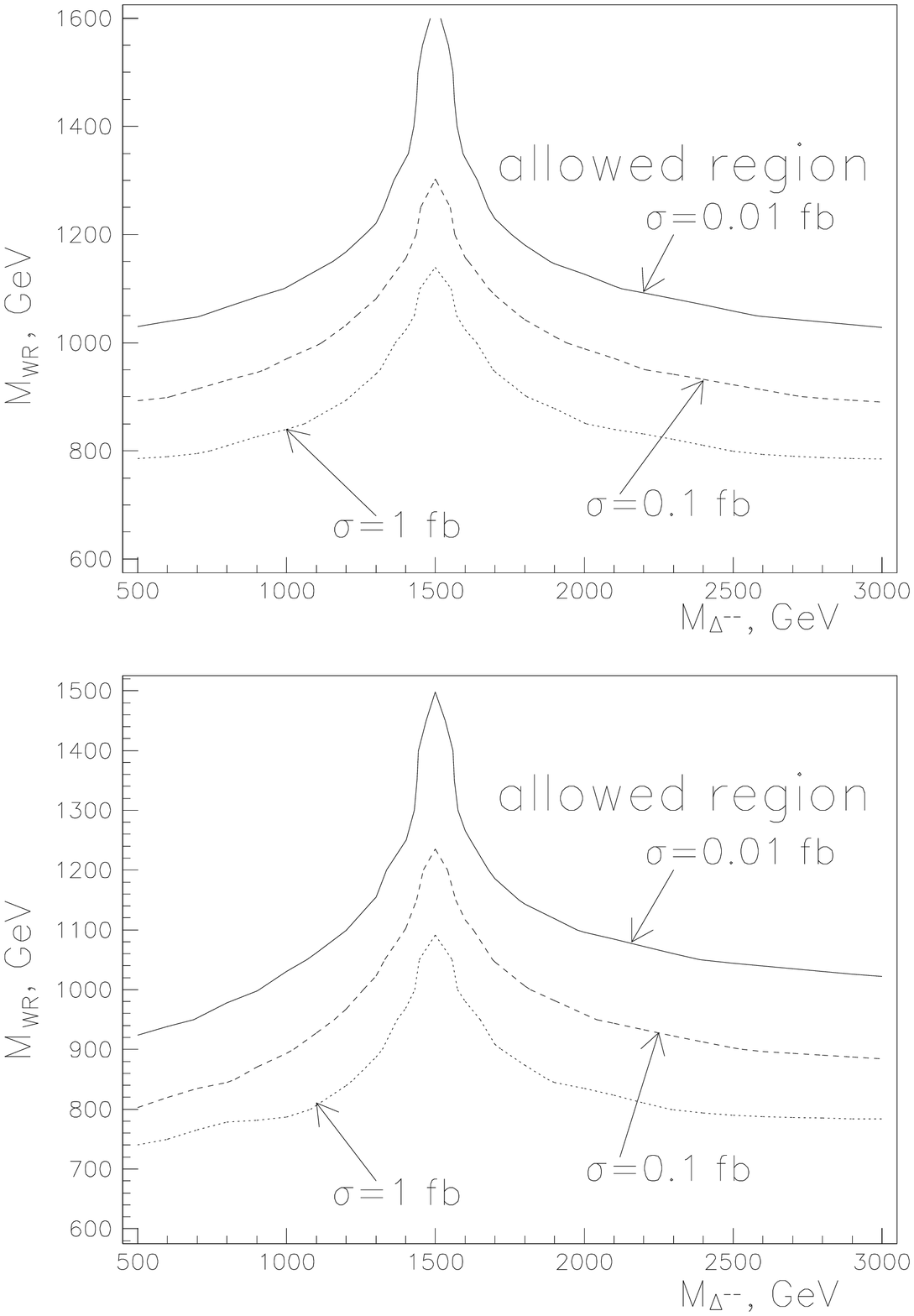} \\
\hspace*{2cm} {\bf (a)}   \hspace*{6cm} {\bf (b)}
 \caption{ \footnotesize Cross section 
for the $e^-e^- \rightarrow b \, b \, \bar{t} \,  \bar{t}$
and it's sensitivity to the masses of $W_R$ and $\Delta_R^{--}$ :
 {\bf (a) }
 $\sigma=0.01 \: $ fb
(30 events per year),
 $\sigma=0.1 \: $ fb
(300 events per year), 
 $\sigma=1 \: $ fb
(3000 events per year)  
for the energy $E=1.5 \:$ TeV , 
and the right-handed neutrino mass
$m_{\nu_2}=1.5 \: $ TeV;
 {\bf (b)}
 $\sigma=0.01 \; $ fb
(30 events per year),
 $\sigma=0.1 \; $ fb
(300 events per year), 
 $\sigma=1 \; $ fb
(3000 events per year)
for the processes with 1 $b$-jet
or with light-quarks only in the final state
(see comments in the text)  
for the energy $E=1.5 \;$ TeV, 
and the right-handed neutrino masses:
$m_{\nu_2}=1.5 \; $ TeV  (on the top)
and
$m_{\nu_2}=1 \; $ TeV (in the bottom).}
\label{s15n15}
\end{figure}

The SM background can be suppressed
to the level 4 orders of magnitude below the
process rate if the proper
cuts in the phase space are
applied, and it can be made even 7 orders of magnitude below the signal level    if the full energy of the event can be
reconstructed with the  accuracy of 50 GeV.   

From figs. 1 and 2 one can see
 that the reaction $e^-e^- \rightarrow q\:q\:\bar{q}\: \bar{q}$
may be observed at LC for a wide range of reasonable parameter values of the
LRM and already
below the $W_R $ threshold.
For the collision energy  $\sqrt s=1.5$ TeV
and luminosity $ 10 ^{35} {\rm cm}^{-2}
\cdot {\rm s}^{-1}$ the lower limit for the mass of
the right-handed gauge boson one could reach is $M_{W_R} \sim 1$ TeV.
More detailed study of this process one can find in \cite{MaalRom1}.

\subsection{Process \prmnqq  as a test of the $\Delta_L$ Higgs.}

The interactions of the $\Delta_L$ field described above are the same for both the SM with
additional Higgs triplet and for the LR-model. The analysis, presented below is therefore valid for both of these models.

In \cite{Rizzo} the production of the singly charged Higgses in $e^-e^-$ collisions
was assumed to take place in pairs through a $W^-W^-$ fusion. This
process conserves the lepton number. We will consider here production
 processes
 that probe the lepton number violating Yukawa couplings. The pair production, which proceeds through t-channel exchange of
Majorana neutrinos and s-channel exchange of $\Delta_L^{--}$,
 is
not a suitable process to study  in this case.
 This is because
  the neutrino exchange is proportional to Majorana mass
 of the neutrino and hence is suppressed and the
 $\Delta_L^{--}\Delta_L^-\Delta_L^- $ vertex depends on the
 self-couplings of scalar potential whose values are unknown. We consider instead
a production of a single $\Delta_L^-$ in the process $e^-e^-\to\Delta_L^- W^-_{\mu}$
where  the
t-channel   neutrino  exchange is not suppressed as the t-channel neutrino
  has the same chirality in the both vertices and  in the s-channel process
the strength of the
 $\Delta_L^{--}\Delta_L^-W_{\mu}^-$ vertex does
  not depend on any unknown parameter of the scalar potential
but is determined by the  gauge coupling. The experimentally
clearest final state to study is the one where
  $ \Delta_L^-$ decays to a muon and a muonic neutrino and  $W^-_{\mu}$
decays into two quark jets (e.g. $d$ and $\bar{u}$) without missing energy.

 In our calculations, made using the CompHEP package \cite{CompHEP},
 we have imposed the following cuts for the final
 state phase space:
 each final state particle has energy greater than 10 GeV
(including neutrino);
 the transverse energy of each particle (including missing
transverse energy) should be greater than 5 GeV;
the opening angle between two quark jets should be more than
$20^o$;
each final state particle should have the outgoing
direction more then $10^o$ away from the beam axis.

In Fig. 3 we present the  dependence of the cross section
of the process \prmnud on the collision energy for the different values of
masses of singly ($M_{\Del}$=100, 400, 700, 1000 GeV)
and doubly charged ($M_{\DDel}=$ 100, 400, 700, 1000 GeV) triplet Higgses.
 The cross sections are dominated by
 the resonance  at $\sqrt{s}=M_{\DDel}$.
  To estimate the width of the peak we have chosen
$\Gamma_{\DDel}=10^{-3}  M$  for the two lepton decays and
the $\DDel \rightarrow \Del W_L^-$ mode was also taken into
account \cite{hRll}.   One may conclude that at 0.01 fb level the
process \prmnud may be observed away from the $\DDel$ resonance and
even below the $\Del$ threshold.

 Fig. 4 presents the sensitivity of the reaction \prmnud on the masses of the
 triplet Higgs particles $\Del$ and $\DDel$
 for the collision energy $ \sqrt{s}=500$
 GeV.  We have estimated the values of  the running coupling constants at 500 GeV
 by applying the approximate RG equations of the SM  \cite{Rom}. The influence
of the triplet Higgses on the running, which can be expected to be
quite small, is not taken into account. \mbox{Fig. 4a} displays
the cross section of the \prmnud process for the different values of
the $\Del$ and $\DDel$ masses, with assuming for the Yukawa
couplings their maximal allowed values that are in accordance with
the present phenomenological constraints \cite{hRll}: $$
h^2_{ee} < 10^{-5}\cdot M_{\DDel} \; {\rm GeV}, $$
\be
h^2_{\mu \mu} < 10^{-5} \cdot M_{\DDel} \; {\rm GeV.}
 \ee
 If the
mass of doubly charged Higgs is considered to be greater than 100
GeV , then $h_{e e} \cdot h_{\mu \mu} < $ 0.18 or $\sqrt{h_{e e}
\cdot h_{\mu \mu}} <$ 0.4.

In
Fig. 4b we show the dependence of the cross section on the $\DDel$
mass in the case that  $\Del$ is effectively decoupled.
Supposing that the mass of $\DDel$ is known, one can conservatively
estimate, by setting for the Yukawa coupling $h_{\mu\mu}$ the
largest phenomenologically allowed value, the contribution of the
$\DDel$ mediated processes on the total cross section. When this
is subtracted from the total cross section, what is left is the
contribution of the t-channel neutrino exchange process alone.
This has  a threshold behaviour and its strength gives direct
information on the product ${h_{e e} \cdot h_{\mu \mu}}$ of Yukawa
couplings.

In Fig. 4c we display the 0.03 fb (30 events per year) discovery
contours on the  $\left( M_{\Del},M_{\DDel} \right)-$plane,
 corresponding to the  cross section of the
  isolated t-channel process, for the
different values (0.1, 0.4 and 1.0) of "average" Yukawa couplings
$(h_{\rm Yuk}=\sqrt{h_{e e} \cdot h_{\mu \mu}})$. In the plot the
collision energy is taken as $\sqrt{s}= 500$ GeV.
 It is seen from the figure that the process \prmnud might
  probe the the mass $M_{\Del}$
  to much larger values than what is
   the production threshold, providing
   that the average Yukawa coupling is
   larger than 0.1 and the collision
     does not happen in the vicinity
     of the $\DDel$ pole. If these
      conditions are not met $\Del$ would
       have detectable effects only when
        it is produced as a real particle.

\begin{figure}[t]
\vspace*{-0.5cm} \hspace*{0.2cm} \epsfysize=6.5cm \epsfxsize=6.5cm
\epsffile{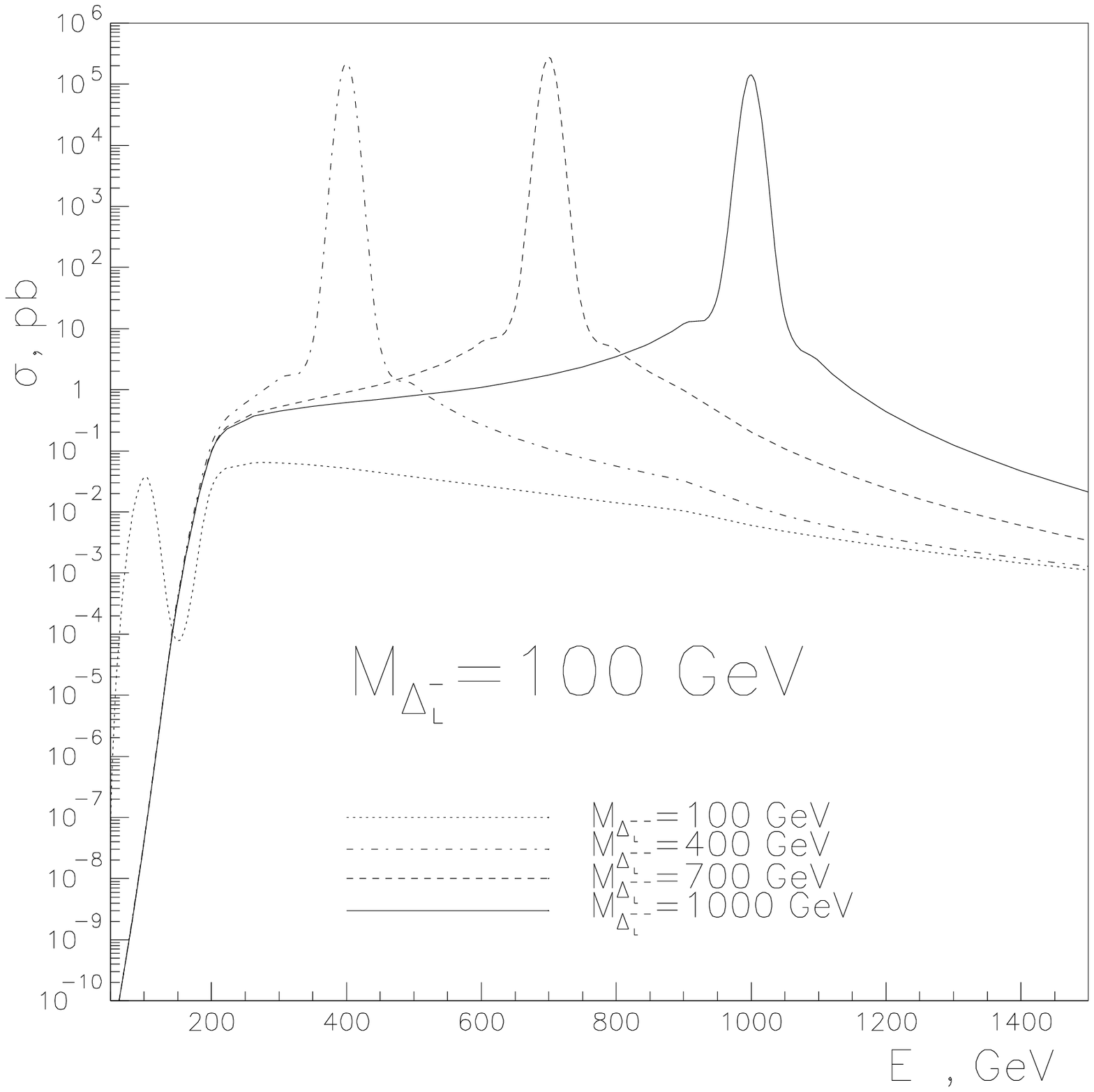} \epsfysize=6.5cm \epsfxsize=6.5cm
\epsffile{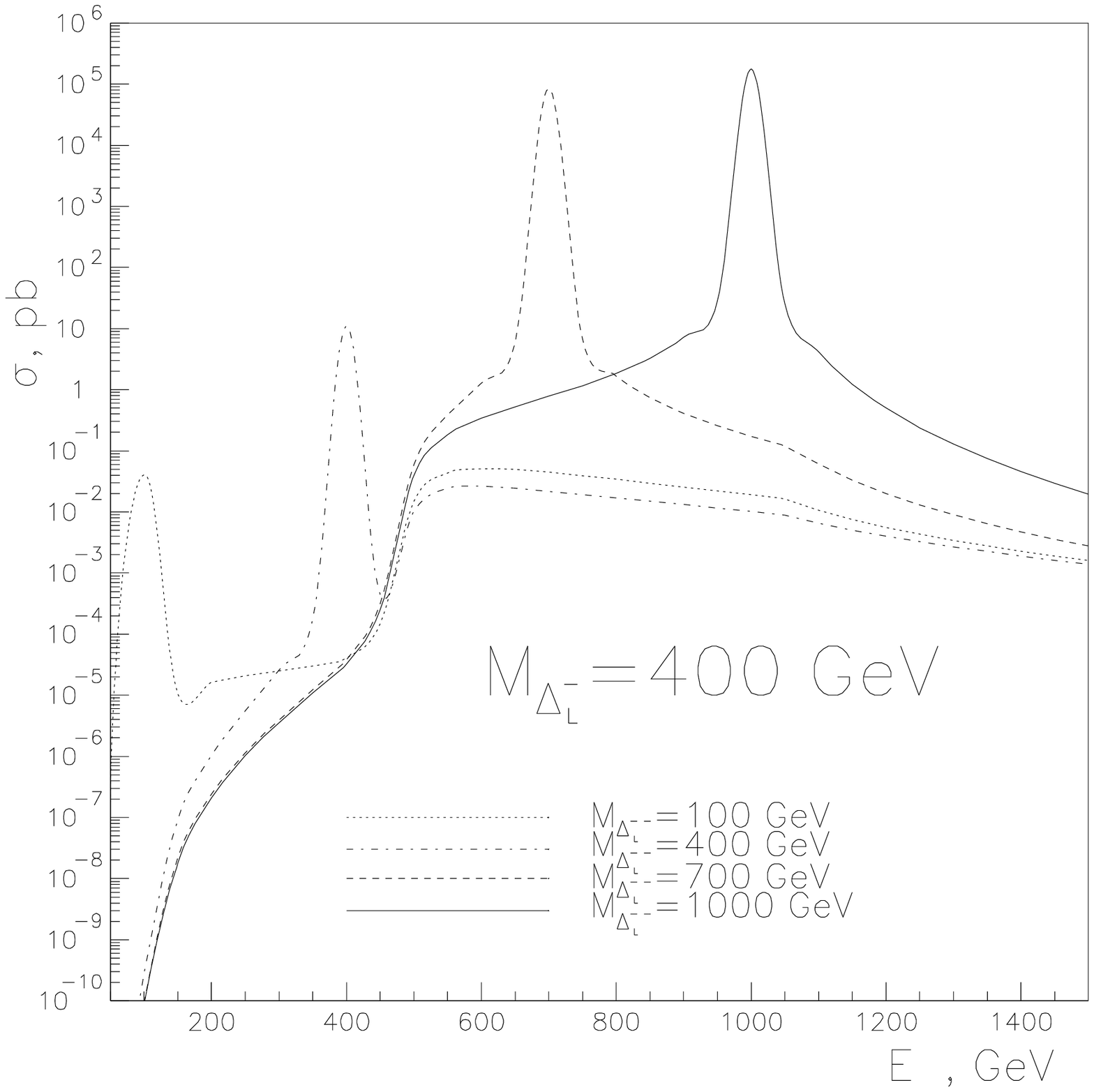} \\ \epsfysize=6.5cm \epsfxsize=6.5cm
\epsffile{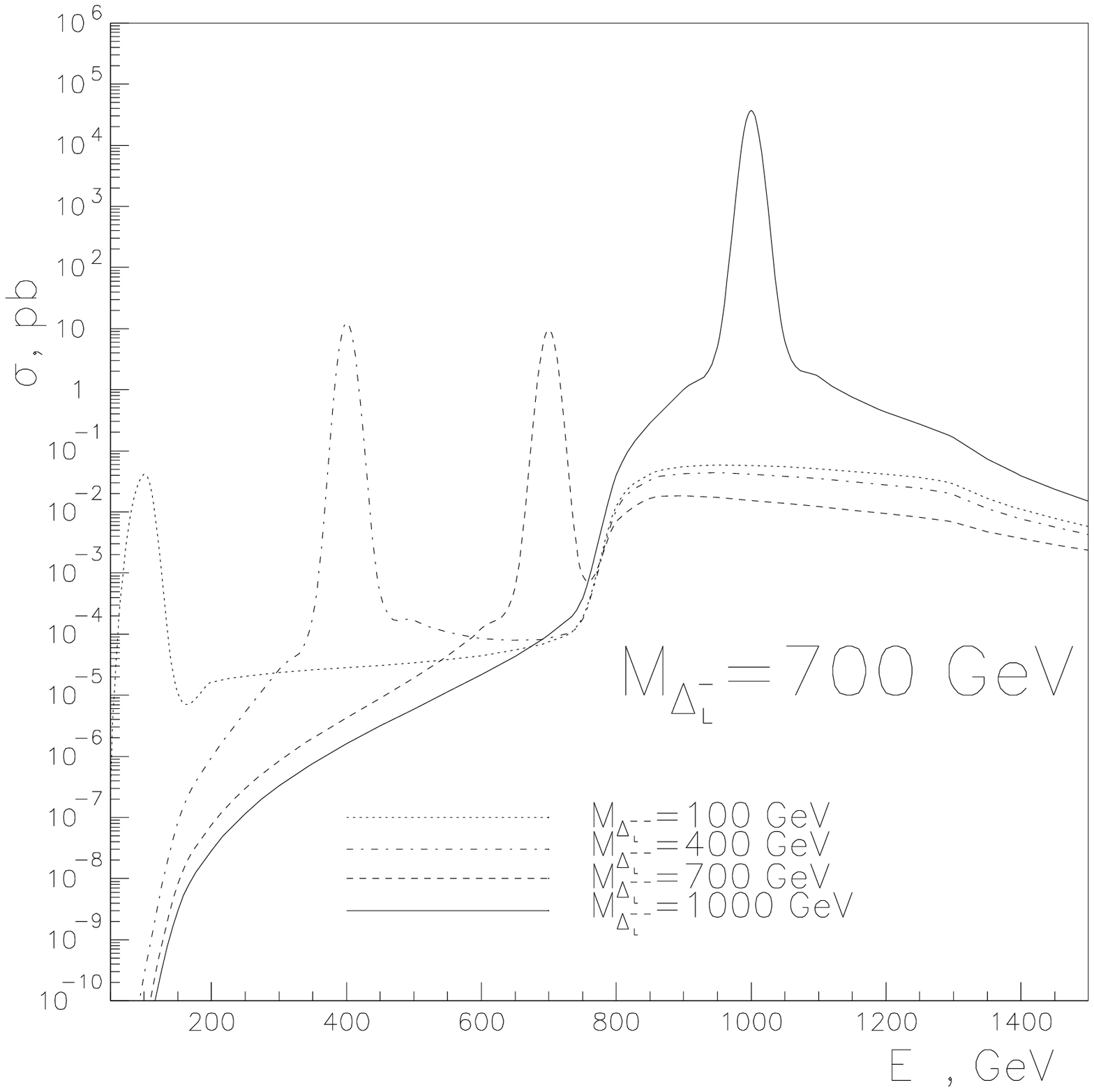} \epsfysize=6.5cm \epsfxsize=6.5cm
\epsffile{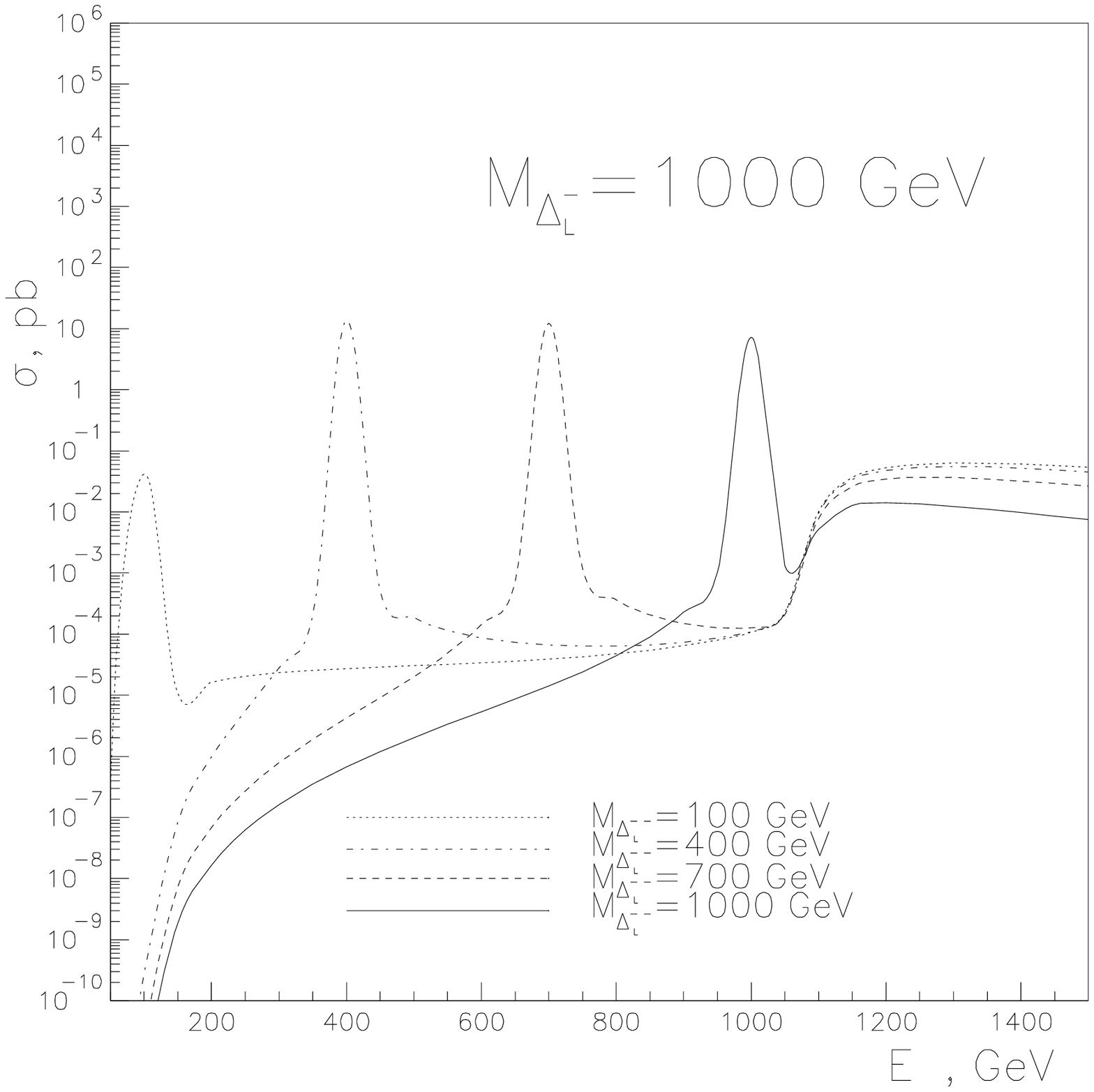} \caption{\footnotesize Energy dependence of the cross
section of the $e^-e^- \rightarrow \mu \, \nu_{\mu} \, d\,
\bar{u}$
 for  different
 values of the masses of singly charged ($\Delta_L^-$) and doubly
charged ($\Delta_L^{--}$)
 triplet Higgses.}
\end{figure}

\begin{figure}[t]
\vspace*{-0.5cm} \hspace*{0.2cm}
\epsfysize=6.5cm \epsfxsize=6.5cm
\epsffile{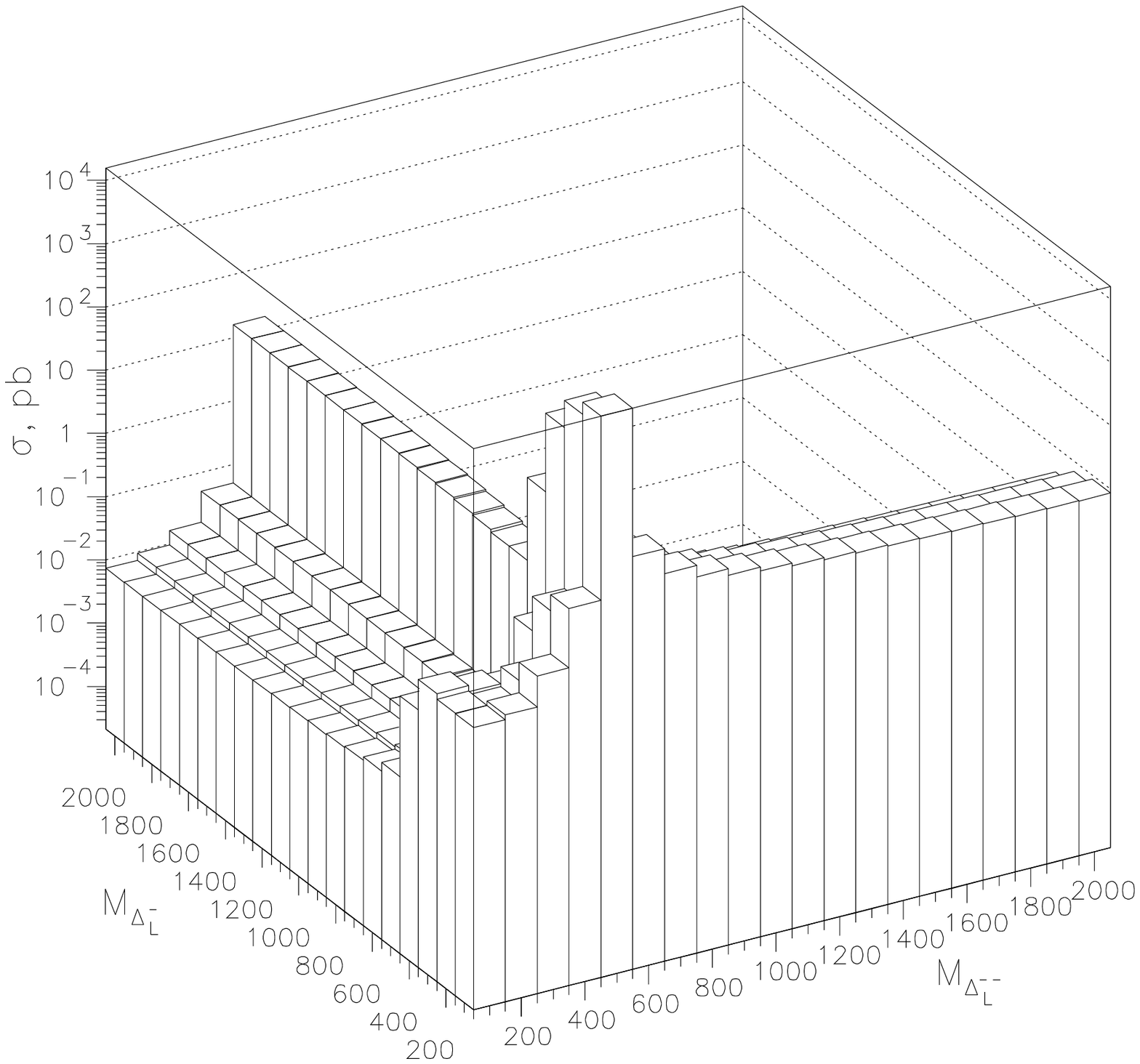}
 \epsfysize=6.5cm \epsfxsize=6.5cm
\epsffile{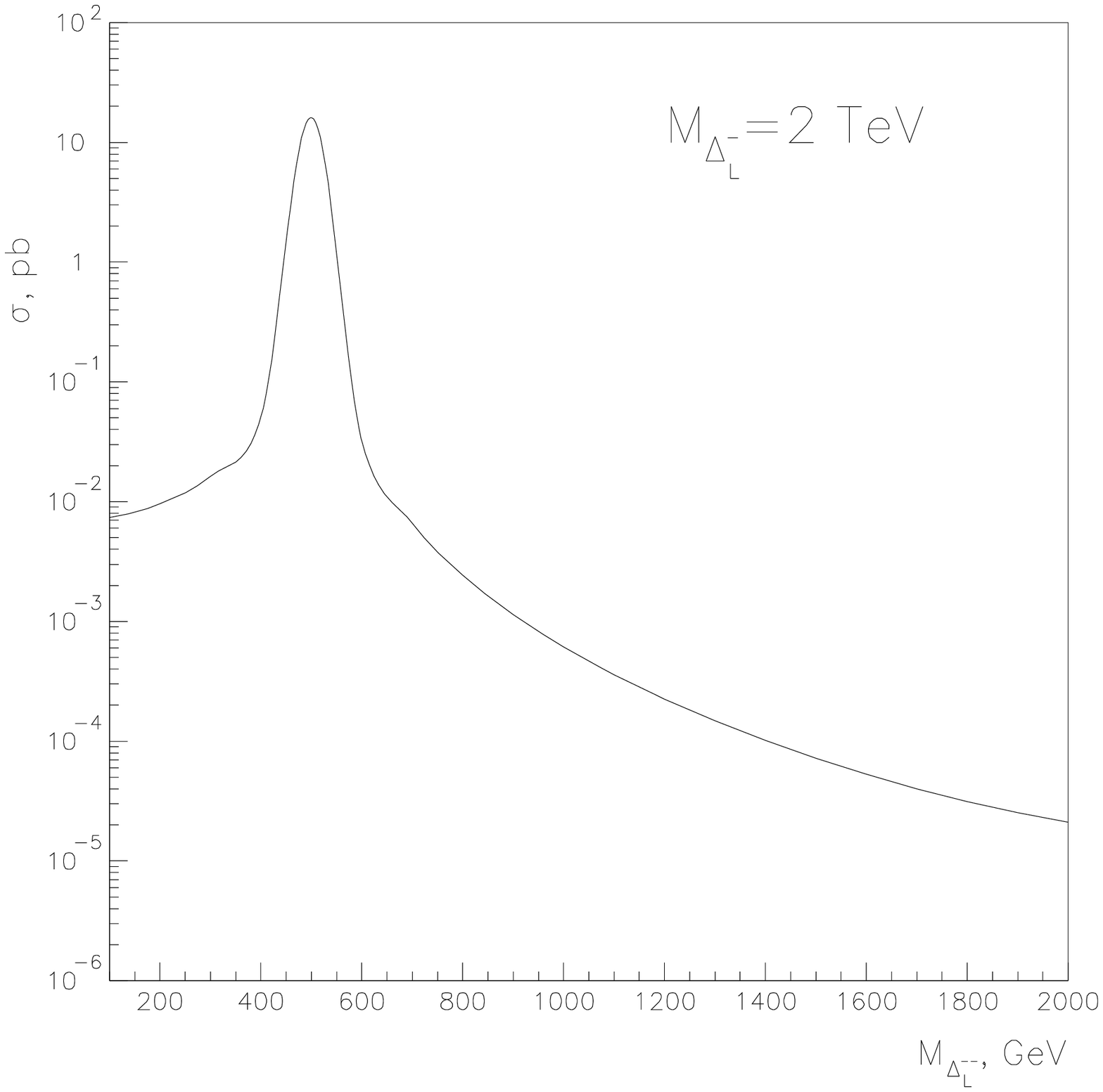} \\
\hspace*{2.7cm} (a) \hspace*{6.7cm} (b) \\
 \hspace*{4cm}
 \epsfysize=6.5cm
\epsfxsize=6.5cm \epsffile{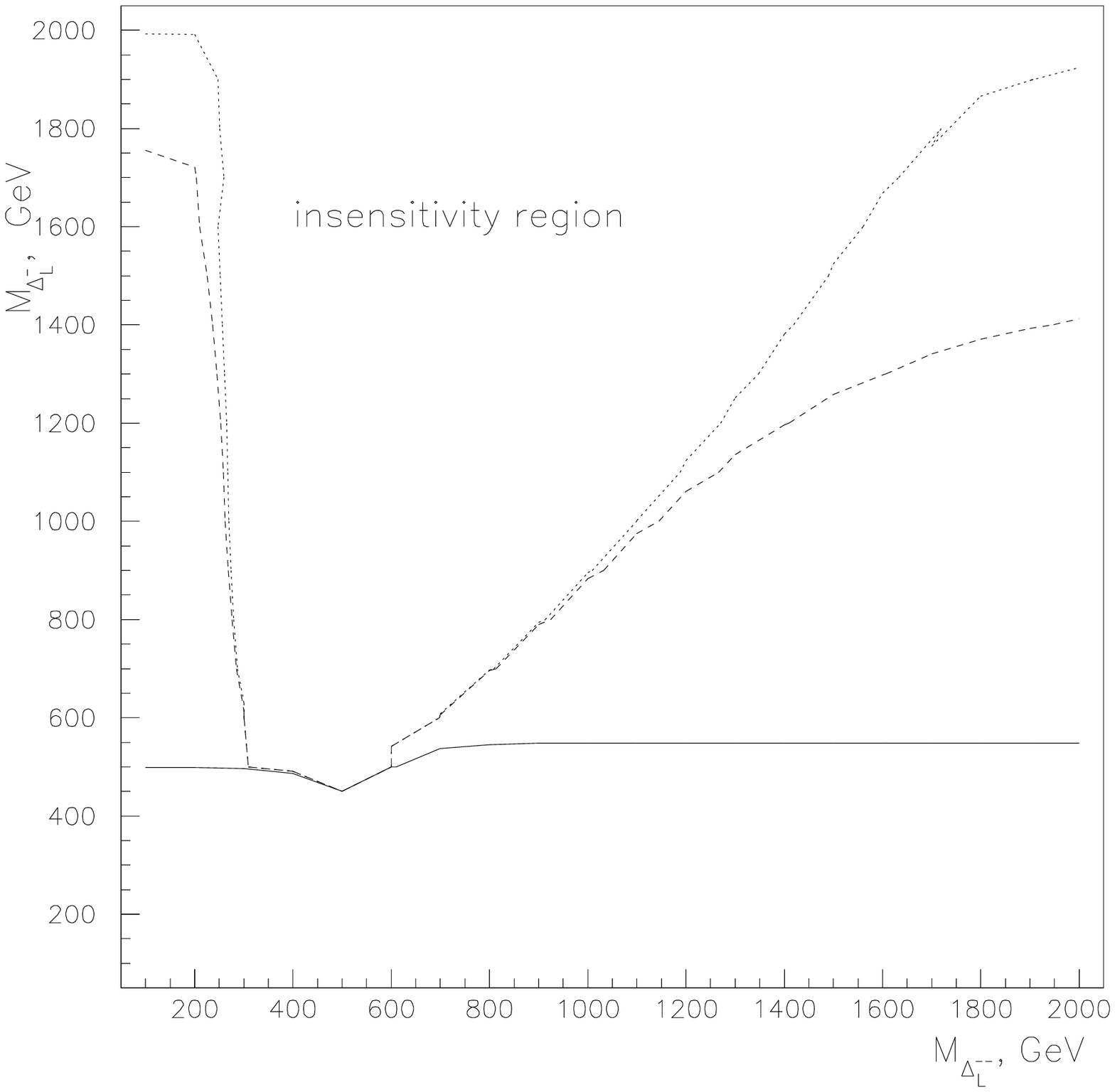} \\
\hspace*{7.1cm} (c)
\caption{ \footnotesize The cross section for the $e^-e^- \rightarrow \mu \,
\nu_{\mu} \, d \, \bar{u}$ process (a)
 for the different
 values of $M_{\Delta^{-}_L}$  and $M_{\Delta^{--}_L}$,
 (b) as a function of  $M_{\Delta^{--}_L}$
 in the limit  $M_{\Delta^{-}_L} > > M_{\Delta^{--}_L}$.
  (c) The contour plots for the difference between cross
sections of the process with finite and infinite
$M_{\Delta^{-}_L}$ for 0.03 fb (30 events per year), for different
values of Yukawa couplings
 (solid line for $h=0.1$, dashed line for $h=0.4$, and dotted line for
 $h=1$).
 Collision energy is taken to be $\sqrt{s}= 500$ GeV.
 In Fig. (c) the region above curves is outside the reach of
 experiment.}
\end{figure}

  The main SM background to the reaction \prmnud is due to the process $e^-e^- \rightarrow W^-W^- \nu
  \nu$ studied in \cite{Cuy}.
 Reconstructing the
   invariant squared mass of the muon and neutrino pair it would
   be possible to separate background in the cases when the
   mass difference between $\Del$ and $W^-$
is greater than  invariant mass resolution (for $M_{\Del} > 100$
GeV this should be possible). But even in the cases when $M_{\Del}
\simeq M_W$ it is possible to compare the cross sections of \prmnud
and $e^-e^- \rightarrow d \: \bar{u} \: s \: \bar{c}$ which should
be equal in the SM. Any substantial  difference between these
cross sections would  be a signal of the new physics. In other
words, in order to get rid of the SM background one should
consider the ratio of the cross sections of $e^-e^- \rightarrow d
\; \bar{u} \; s \; \bar{c}$ and \prmnud.

From figs 3 and 4 one can conclude that the process \prmnud provides a good test for
 lepton flavor non-conservation of the singly charged scalars. At the
 collision energy 500 GeV the process may be seen well below
 $\Del$ and/or $\DDel$ thresholds for a wide range of the lepton number violating Yukawa
 couplings. The influence of $\Del$ contribution
 (below its threshold) may be
 extracted from the process, if colliding energy
 is away from the  $\DDel$  resonance. The present bounds on the
 Yukawa couplings may be significantly improved.
More detailed study of this process one can find in \cite{MaalRom2}.

\section{The supersymmetric left-right models\label{sec:models}}

The  supersymmetrized versions of the left-right model
\cite{Cvetic:1984su}-\cite{HPP3}
have been actively studied in recent years.
From the supersymmetric point of view, an important motivation for the
left-right models is due to their gauge group.
It has been noted 
\cite{Mohapatra:1986su,Martin:1992mq} that if the gauge symmetry
of MSSM is suitably extended to contain $U(1)_{B-L}$, the 
R-parity is automatically conserved in the Lagrangian of the theory.
Thus one of the major problematic features of the MSSM is explained by
the gauge symmetry.
Here we will concentrate on the supersymmetric left-right model (SLRM)
based on the gauge group 
$SU(3)_C\times SU(2)_L \times SU(2)_R\times U(1)_{B-L}$ .

The particle content of the supersymmetric left-right model is
enlarged compared to the MSSM.
Because of the extended symmetry, there are superfields containing
gauge bosons connected to $SU(2)_R$.
Due to the more symmetric treatment of the fermions in the model, also
the right-handed neutrino superfield is included.
The breaking of the extended symmetry requires a new set of Higgs 
bosons.
The Higgs sector of the SLRM can be chosen in many ways, but with
triplets in the spectrum, one can have the conventional see-saw mechanism for
neutrino mass generation.
This will be the choice for the $SU(2)_R$ breaking here. 
The $SU(2)_L$ will be broken mainly by bidoublets which contain the
doublets of the MSSM:
\begin{eqnarray}
&&\Phi = \left( \begin{array}{cc} \Phi^0_1 & \Phi^+_1 \\ \Phi^-_2 &
\Phi^0_2 \end{array} \right),\;  \chi = \left(
\begin{array}{cc} \chi^0_1 & \chi^+_1 \\ \chi^-_2 & \chi^0_2
\end{array} \right) \sim (1,2,2,0) , \nonumber \\ 
&&\Delta_R  
\sim (1,1,3,-2),\;  \delta_R 
\sim (1,1,3,2), \; \Delta_L 
\sim (1,3,1,-2),\;  \delta_L   \sim
(1,3,1,2) .
\label{eq:fields2}
\end{eqnarray}
In (\ref{eq:fields2}) it is assumed that the gauge symmetry is
supplemented by a discrete left-right symmetry.
The $SU(2)_L$ triplets $\Delta_L$ and $\delta_L$  
make the Lagrangian fully symmetric under $L \leftrightarrow
R$ transformation, although these are not needed
for symmetry breaking, or for the see-saw mechanism.

The most general gauge invariant superpotential involving these
superfields can be written as (generation indices suppressed)
\begin{eqnarray}
W_{min}&=& h_{\Phi Q}Q^T i\tau_2 \Phi Q^c + h_{\chi Q}Q^T i\tau_2 \chi Q^c +
h_{\Phi L}L^T i\tau_2 \Phi L^c
+ h_{\chi L}L^T i\tau_2 \chi L^c
 \nonumber\\ &&
+h_{\delta_L} L^T i\tau_2 \delta_L L + h_{\Delta_R}
L^{cT} i\tau_2 \Delta_R L^c
+ \mu_1 \rm{Tr} (i\tau_2\Phi^T i\tau_2
\chi) + \mu_1' \rm{Tr} (i\tau_2\Phi^T i\tau_2 \Phi) 
 \nonumber\\ &&
+\mu_1'' \rm{Tr} (i\tau_2\chi^T i\tau_2 \chi) +\rm{Tr}
(\mu_{2L}\Delta_L \delta_L + \mu_{2R}\Delta_R\delta_R).
\label{eq:superpotential}
\end{eqnarray}
The general form of the vacuum expectation values of the various
scalar fields which preserve the $U(1)_{\rm{em}}$ 
gauge invariance can be written as
\begin{eqnarray}
&&\langle \Phi \rangle = \left( \begin{array}{cc} \kappa_1 & 0 \\ 0 &
e^{i \phi_1}\kappa_1' \end{array} \right) ,\;  
\langle \chi \rangle =\left(\begin{array}{cc} e^{i \phi_2} \kappa_2' &
0 \\ 0 & \kappa_2\end{array} \right) , \;\nonumber \\ 
&& 
\langle \Delta_R^0 \rangle = v_{\Delta_R} ,\;  
\langle \delta_R^0 \rangle = v_{\delta_R} ,\;
 \langle \Delta_L^0 \rangle =v_{\Delta_L}  ,\;  
\langle \delta_L^0 \rangle = v_{\delta_L} , \;
\langle \tilde\nu_L \rangle = \sigma_L ,\;  
\langle\tilde\nu_L^c\rangle =  \sigma_R.
\label{eq:vevs}
\end{eqnarray}
{}From the heavy gauge boson mass limits the triplet vacuum
expectation values $v_{\Delta_R}$ and $v_{\delta_R}$ are in the range 
$v_{\Delta_R},v_{\delta_R} \gsim 1$ TeV.  
$\kappa_1'$
and $\kappa_2'$ contribute to the mixing of the charged gauge bosons
and to the flavour changing neutral currents, and are usually assumed
to vanish. 
As was pointed out before, in order to have the tree level value of the electroweak $\rho$
parameter close to unity 
the left-triplet vacuum expectation values $v_{\Delta_L}$ and $v_{\delta_L}$
must be small.

With the minimal field content, the only way to preserve the
$U(1)_{em}$ gauge symmetry unbroken is to have a nonzero sneutrino
VEV \cite{Kuchimanchi:1993jg,Huitu:1995zm}.
Thus the R-parity is spontaneously broken in the SLRM with minimal
particle content and renormalizable interactions.

An alternative to the minimal left-right supersymmetric
model described above
involves  additional  triplet fields, 
$\Omega_L(1,3,1,0)$ and $\Omega_R(1,1,3,0)$ 
to the minimal model \cite{Aulakh:1997ba}.
In these extended models the breaking of $SU(2)_R$ is achieved in two
stages via an intermediate  symmetry 
$SU(2)_L \times U(1)_R \times U(1)_{B-L}$. 
In this theory the parity-breaking minimum
respects the electromagnetic gauge invariance without a sneutrino VEV.
The superpotential for these models
contains additional terms involving the triplet fields
$\Omega_L$ and $\Omega_R$: 
\begin{eqnarray}
 W_{\Omega}&=& W_{min} + 
 \frac 12 \mu_{\Omega_L}   {\rm{Tr}}\,\Omega_L^2
	  +\frac 12 \mu_{\Omega_R}  {\rm Tr}\,\Omega_R^{2} +a_L {\rm Tr}\,\Delta_L \Omega_L \delta_L
       +a_R {\rm Tr}\,\Delta_R \Omega_R \delta_R \nonumber \\
& &  + {\rm Tr}\, \Omega_L  \left( \alpha_L \Phi i\tau_2 \chi^T
 i\tau_2+{\alpha_L}' \Phi i\tau_2 \Phi^T i\tau_2 +{\alpha_L}'' \chi i\tau_2
 \chi^T i\tau_2  \right)
 \nonumber \\ & & + {\rm Tr}\, \Omega_R \left( \alpha_R i \tau_2 \Phi^T i\tau_2
 \chi+{\alpha_R}' i\tau_2 \Phi^T i\tau_2 \Phi +{\alpha_R}'' i\tau_2\chi^T i\tau_2
 \chi  \right),
\label{superpot}
\end{eqnarray}
where $W_{min}$ is the superpotential (\ref{eq:superpotential}) of the  minimal
left-right model.
In these models the see-saw mechanism takes its canonical
form with $m_{\nu} \simeq m_D^2/M_{BL}$, where $m_D$
is the neutrino Dirac mass. 
In this case the low-energy effective theory
is the MSSM with unbroken $R$-parity, and contains besides the usual 
MSSM states, a triplet of Higgs scalars much lighter than the 
$B-L$ breaking scale.

Another possibility is to add non-renormalizable terms
to the Lagrangian of the minimal left-right supersymmetric model,
while retaining the minimal Higgs content
\cite{Martin:1992mq,Aulakh:1997ba}.
The superpotential for these models 
can be written as
\begin{eqnarray}
W_{NR} &=& W_{min} +  {a_L \over 2 M} ({\rm Tr}\, \Delta_L \delta_L)^2
 +  {a_R \over 2 M} ({\rm Tr}\, \Delta_R \delta_R)^2+ {c \over M}{\rm Tr}\,
 \Delta_L \delta_L {\rm Tr}\, \Delta_R \delta_R \nonumber \\
& & + {b_L \over 2 M} {\rm Tr}\, \Delta_L^2 {\rm Tr}\, \delta_L^2 +
 {b_R \over 2 M} {\rm Tr}\, \Delta_R^2 {\rm Tr}\, \delta_R^2 + {1 \over M}
 \left[ d_1{\rm Tr}\, \Delta_L^2 {\rm Tr}\,  \delta_R^2 +
d_2 {\rm Tr}\, \delta_L^2 {\rm Tr}\, \Delta_R^2 \right]
  \nonumber \\
 & &
+ {\lambda_{ijkl}\over  M}{\rm Tr}\,i\tau_2 \Phi_i^T i\tau_2
\Phi_j{\rm Tr}\,i\tau_2 \Phi_k^T i\tau_2
\Phi_l + {\alpha_{ijL}\over M} {\rm Tr}\, \Delta_L  \delta_L
\Phi_i i\tau_2 \Phi_j^T i\tau_2   \nonumber\\
&&+  {\alpha_{ijR}\over M}
{\rm Tr}\, \Delta_R  \delta_R  i\tau_2 \Phi_i^T i\tau_2 \Phi_j 
+ {1 \over M}{\rm Tr}\,\tau_2 \Phi_i^T \tau_2 \Phi_j [\beta_{ijL}
{\rm Tr}\, \Delta_L
 \delta_L + \beta_{ijR} {\rm Tr}\, \Delta_R \delta_R ] \nonumber \\
& &+ 
{\eta_{ij} \over M } {\rm Tr}\,\Phi_i  \Delta_R i\tau_2
 \Phi_j^T i \tau_2 \delta_L
+
{\overline \eta_{ij} \over M } {\rm Tr}\,\Phi_i  \delta_R
i\tau_2 \Phi_j^T i\tau_2
 \Delta_L  \nonumber \\
& & + { k_{ql} \over M}Q^T i\tau_2 L\,Q^{cT} i\tau_2 L^c
 + {k_{qq} \over M}Q^T i\tau_2 Q\,Q^{cT} i\tau_2 Q^c
+ { k_{ll} \over M}L^T i\tau_2 L\,L^{cT} i\tau_2 L^c
\nonumber \\ & &+
{1\over M} [j_L Q^T i\tau_2 Q\,Q^T i\tau_2 L +  j_R
Q^{cT} i\tau_2 Q^c \,Q^{cT} i\tau_2 L^c].
\label{nonsuperpot}
\end{eqnarray}
It has been shown that the addition of non-renormalizable
terms suppressed by a high scale such as Planck mass,
$M_{Pl} \sim 10^{19}$ GeV, with the minimal field content
ensures the correct pattern of symmetry breaking
in the supersymmetric left-right model with the intermediate
scale $M_R \gsim 10^{10} - 10^{11}$ GeV, and $R$-parity remains
exact.

In addition to the lightest neutral CP-even Higgs, it has been known
for quite some time \cite{Huitu:1995zm} that the lightest doubly 
charged Higgs boson occurring in triplets of the SLRM may be light.
The detection of a doubly charged Higgs was discussed extensively 
in \cite{123E}.
Also the fermionic sector of the Higgs sector and its use in
identifying the model was considered in \cite{123E}.
While the lightest doubly charged Higgs or its fermionic partner may 
offer best possibilities to identify the triplet, 
the singly charged chargino production may be most suitable for 
finding out the R-parity violation in the model \cite{HMP}.
Here we will first concentrate on the experimentally interesting
results on the masses of the lightest neutral and doubly charged 
Higgs bosons, and then describe new results in analysing different
processes.

\subsection{The upper limit on the lightest CP-even Higgs\label{sec:h0mass}}

In the case of the SLRM we have many
new couplings and also new scales in the model and it is
not obvious, what is the  upper limit on the lightest CP-even Higgs 
boson mass.
This mass bound is a very important issue, 
since the experiments are  approaching
the upper limit of the lightest Higgs boson mass in the MSSM.

A general method to find an upper limit for the lightest Higgs mass  
was presented in \cite{Comelli:1996xg}.
This method has been applied to the mass of the lightest Higgs, 
$m_{h}$, of SLRM \cite{HPP3} in three cases:
(A) $R$-parity is spontaneously broken (sneutrinos get VEVs),
(B) $R$-parity is conserved because of additional triplets, and
(C) $R$-parity is conserved because of nonrenormalizable terms.

For the minimal model, case (A), the upper bound on $m_{h}$ is  
\cite{HPP3}
\bea
m_h^2 \leq \frac 1{2 v^2} \left[ g_L^2 (\omega_\kappa^2+\sigma_L^2)^2+
g_R^2 \omega_\kappa^4+g_{B-L}^2 \sigma_L^4 
+ 8 (h_{\Phi L}
\kappa_1'+h_{\chi L} \kappa_2 )^2 \sigma_L^2 + 8 h_{\Delta_L}^2
\sigma_L^4 \right] ,
\label{eq:treeupper2}
\eea
where
$v^2 = \kappa_1^2+ \kappa_1'^2+\kappa_2^2+\kappa_2'^2+\sigma_L^2$ and
$\omega_\kappa^2 = \kappa_1^2-\kappa_2^2-\kappa_1'^2+\kappa_2'^2 .$
The addition of extra triplets does not change 
this bound.
Thus, the bound for the case (B), can be obtained from
(\ref{eq:treeupper2}) by taking the limit $\sigma_L\rightarrow 0$.
The total number of nonrenormalizable terms in case (C) is
large.
However, the contribution to the Higgs mass bound from these
terms is found to be \cite{HPP3}
typically numerically negligible.
Therefore the upper bound for this class of models is essentially the
same as in the case (B).

The radiative corrections to the lightest Higgs
mass are significant.
For the SLRM lightest Higgs they have been calculated in detail
\cite{HPP3}.
For nearly degenerate stop masses, 
the radiative corrections on $m_h$  in the SLRM 
differ in form from the MSSM upper bound only because of
new supersymmetric Higgs mixing parameters. 
\begin{figure}
\vspace*{-0.5cm} \hspace*{0.2cm}
\epsfxsize=79mm
\begin{center}
\mbox{\epsffile{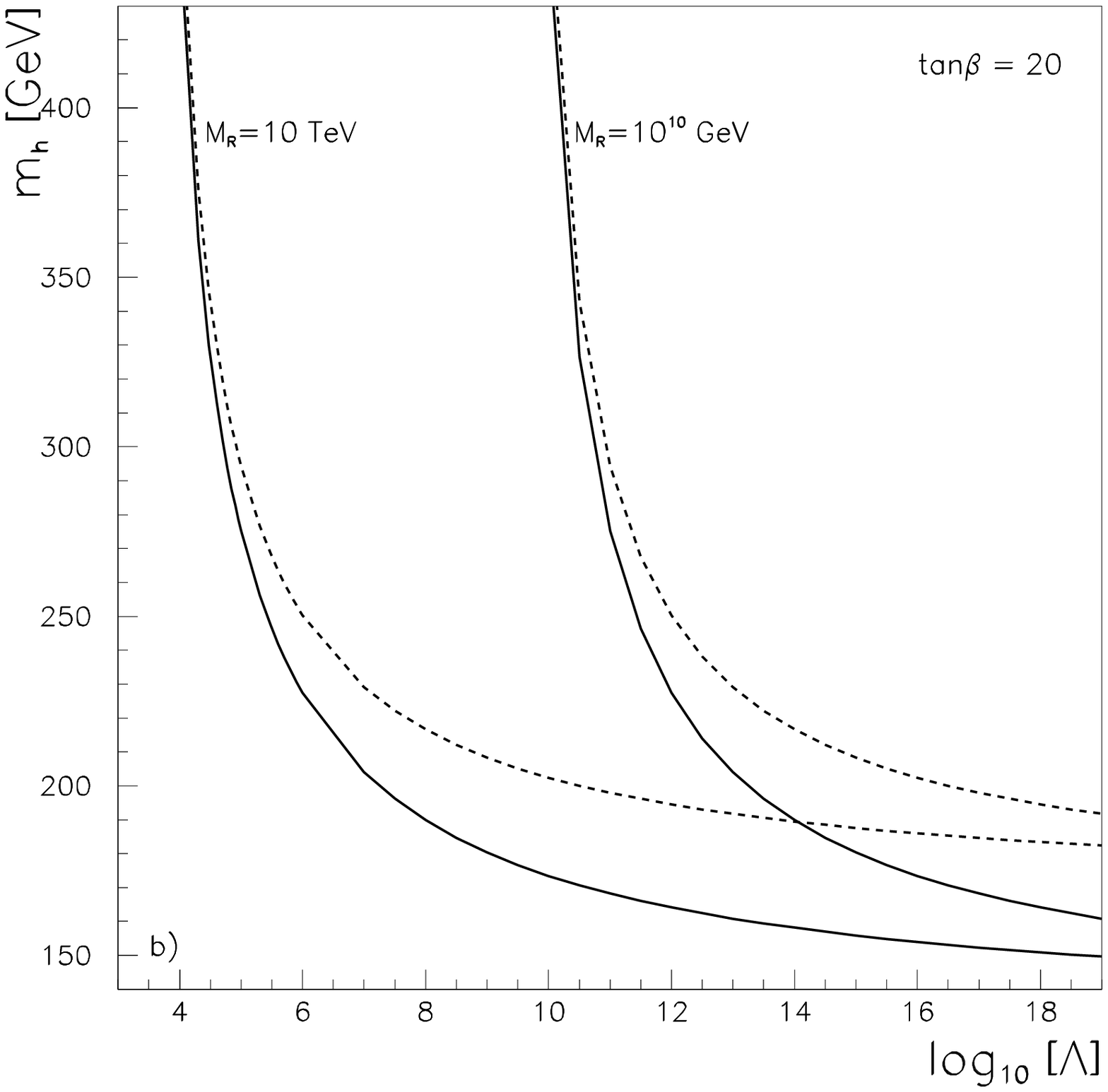}}
\end{center}
\captive{The upper bound on the mass 
of the lightest neutral Higgs boson.
The soft supersymmetry breaking parameters are  1 TeV
(solid) and 10 TeV (dashed).
\label{mh1}}
\end{figure}

The upper bound on 
the mass of the lightest Higgs is plotted in Fig.\ref{mh1} 
as a function of the scale $\Lambda $ up to
which the SLRM remains
perturbative.  
The upper bound is shown for two different values of
the $SU(2)_R$ breaking
scale, $M_R=10$ TeV and $M_R=10^{10}$ GeV,
and for two values of soft supersymmetry breaking
mass parameter, $M_s=1$ TeV and $M_s=10$ TeV.
For large values of $\Lambda $ the upper bound 
is below 200 GeV.

Another relevant issue concerning the lightest Higgs is its branching
ratios to fermions.
Connected to that in the left-right symmetric models are problems with
FCNC, which are expected if several light Higgs bosons exist \cite{EGN},
unless  $m_{H_{FCNC}}\gsim {\cal{O}}(1$ TeV). 
Thus the relevant limit to discuss is the one in which all the neutral Higgs
bosons, except the lightest one, are heavy.
It has been shown that in the decoupling limit the 
Yukawa couplings of the $\tau$'s are the same in the SM and the SLRM 
even if the $\tau$'s contain a large fraction of gauginos or
higgsinos \cite{HPP3}.
If a neutral Higgs boson, which couples to fermions very differently
than the Standard Model Higgs, is found, the model most probably is
not left-right symmetric.

\subsection{The lightest doubly charged Higgs\label{sec:hpp}}

Whether the lightest doubly charged Higgs is observable in experiments 
is an interesting issue, since this particle may both reveal the nature
of the gauge group and help to determine the particular supersymmetric
left-right model in question.
The chances for detection depend strongly on the mass of the particle.
This will be our main concern in this section, but we'll also shortly
review the processes discussed more thoroughly in \cite{123E}.

There are four doubly charged Higgs bosons in the SLRM, of which
two are right-handed and two left-handed.
The masses of the left-handed triplets are expected to be of the same 
order as the soft terms.
The mass matrix for the right-handed triplets depends on the
right-triplet VEV.
Nevertheless,
it was noticed in \cite{Huitu:1995zm} that in the SLRM with broken R-parity
one right-handed doubly charged scalar tends to be light.
Also,  in the nonrenormalizable case
it is possible to have light doubly charged scalars \cite{CM58}.
On the other hand, in the nonsupersymmetric left-right model all the
doubly charged scalars typically have a mass of the order of the 
right-handed scale \cite{GGMKO}.
This is also true in the SLRM with enlarged particle content
\cite{Aulakh:1997ba}.
Thus a light doubly charged Higgs would be a strong indication of a
supersymmetric left-right model with minimal particle content.

\begin{figure}
\vspace*{-0.5cm} \hspace*{0.2cm}
\epsfxsize=68mm
\epsffile{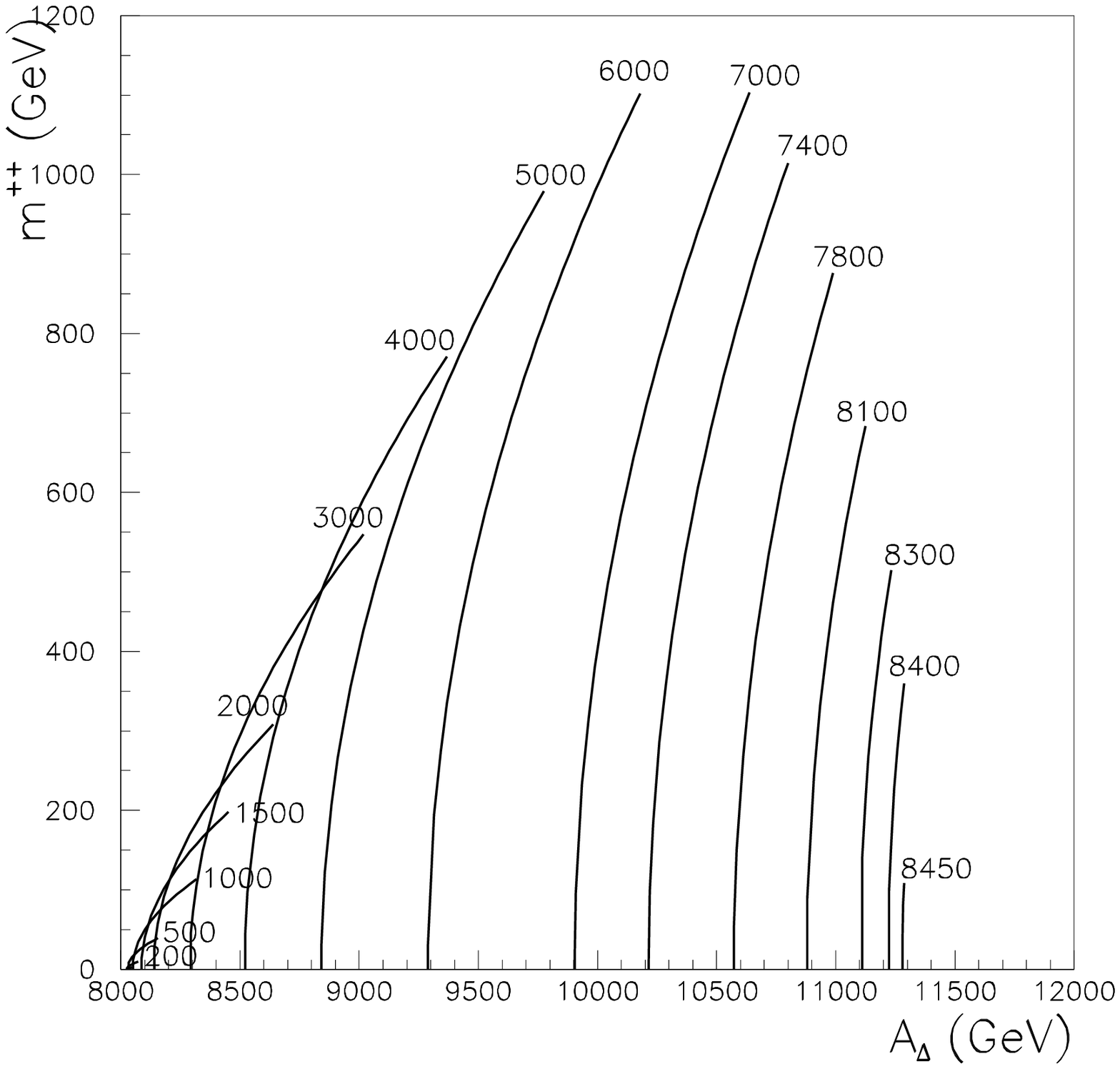}
\epsfxsize=68mm
\epsffile{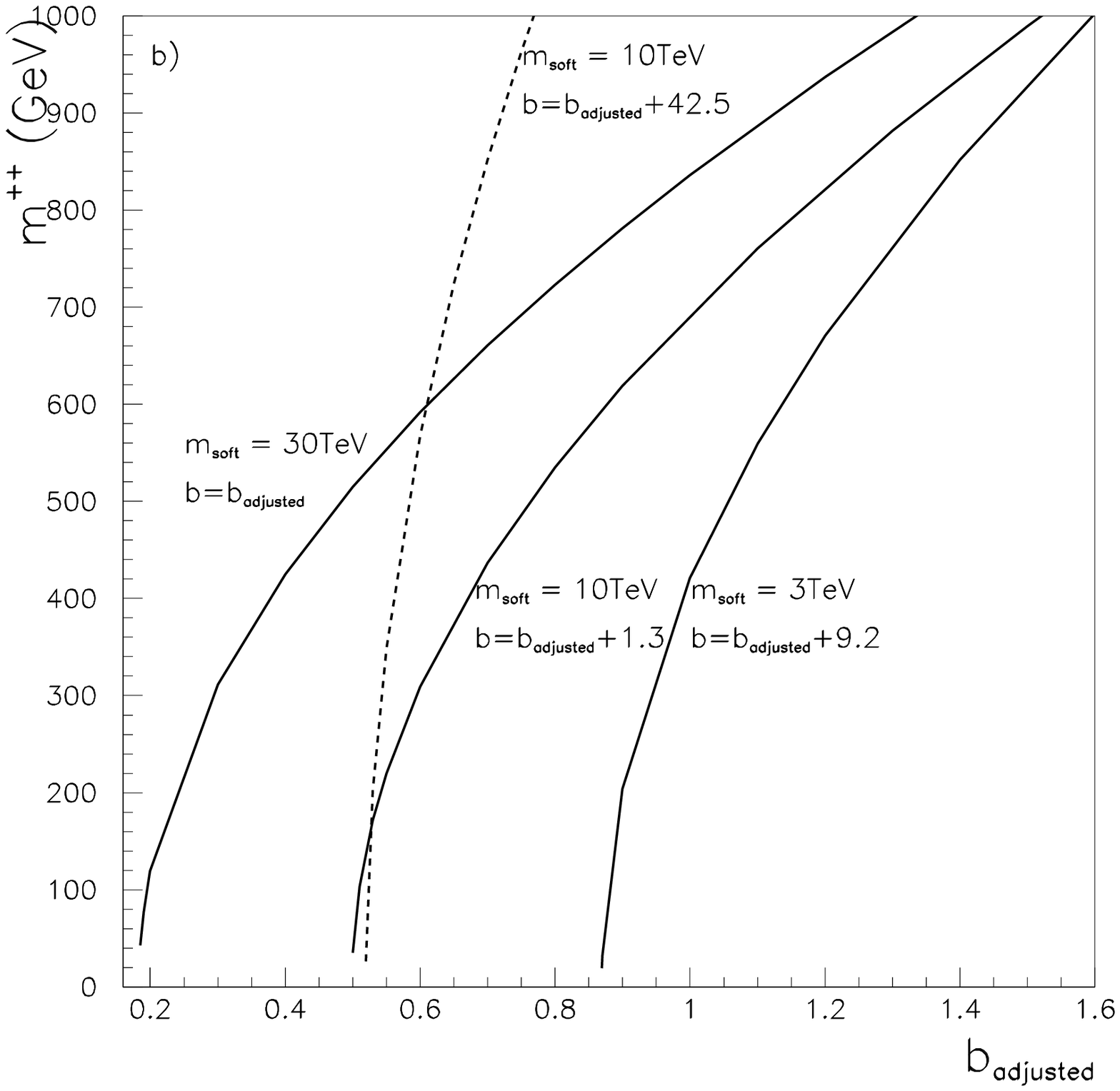} \\
\caption{The mass $m_{H^{++}}$ of the lightest doubly charged Higgs.
In a) the mass is as a function of the soft trilinear coupling $A_\Delta$.
The allowed $\sigma_R$ varies between 100 GeV and 8.45 TeV.
In b) the mass is as a function of the
$b_{adjusted}$-parameter related to $b_R$ as denoted.
In b) $D =(3$ TeV)$^2$ (solid) except  for 
$m_{soft} =10$ TeV also $D=10$ TeV$^2$ is shown (dashed).
The soft supersymmetry breaking parameters
are marked in the figure.
$\tan\beta=50$.
\label{mpp_ad}}
\end{figure}

In Figure \ref{mpp_ad} a) an example of $H^{++}$ masses with broken
R-parity is shown as a function of $A_\Delta $ for fixed 
$\sigma_R$.  The soft masses
and right-handed breaking scale,
are of the order of 10 TeV.
The maximum triplet Yukawa coupling allowed by positivity of the mass
eigenvalues in this case is 
$h_\Delta\sim 0.4$.
Even in the maximal case the mass of the
doubly charged scalar $m_{H^{++}}\sim 1 $ TeV.
In Fig. \ref{mpp_ad} b) $m_{H^{++}}$ is plotted  
in the model containing nonrenormalizable terms 
as a function of the nonrenormalizable $b_R$-parameter for 
$v_R^2/M=10^2$ GeV.

The collider phenomenology of the doubly charged scalars has
been actively studied, since they appear in several extensions of the
Standard Model, can be relatively light and have clear signatures.
The main decay modes for relatively light doubly charged Higgs
are \cite{hRll}
$H^{--}\rightarrow l_1^-l^-_2$, where $l_{1,2}$ denote leptons.  
Thus the experimental signature of the decay is
a same sign lepton pair with no missing energy, including 
lepton number violating final states.

Since the left-right models contain many extra parameters
when compared with the MSSM, a great advantage of the pair production
is that it is relatively model independent.
The doubly charged Higgses can be produced in 
$f\bar f\rightarrow\gamma^*,Z^*\rightarrow H^{++}H^{--}$ both at
lepton and hadron colliders, if kinematically allowed,
even if $W_R$ is very heavy,  
or the triplet Yukawa couplings are very small.
The pair production cross section at a linear collider has been 
given in \cite{GMS}.
The cross section remains sufficiently large close to the kinematical
limit for the detection to be possible.

Kinematically,  production of a single doubly charged scalar would be
favoured.
This option is more model dependent, but for reasonable parameter
range the kinematical reach is
approximately doubled compared to the pair production.

\section{Acknowledgements}
This work has been  supported also by the Academy of Finland under the project no.
 40677 and no 163394, and by RFFI grant 98-02-18137.

\end{document}